\documentstyle[aps,epsfig,multicol]{revtex}
\begin{document} \title{A  Model of Quantum Field Theory with Inter Source}
\author{Gang Zhao \\ Department of Physics, Texas A\&M University, College Station, TX 77840}
\maketitle
\address{Department of Physics, Texas A&M University, College Station, TX 77840}
\begin{abstract}
By putting a confined inter source, we construct a model which can
give us convergent solution from free field equation. On the other
hand, the solution of new field equation can be separated into two
parts, one part is just same as the one in Quantum Field Theory
and make it survived in this model, and the other part, which we
will see doesn't take energy and momentum, just gives us a
negative propagator which can soften quadratic divergence.

\medskip \noindent e-mail: gzhao@physics.tamu.edu
\end{abstract}
\section{Introduction}
It is impossible for us to find convergent stationary solution
from free Schr\"{o}dinger equation or free field equation. But if
we put an inter source which moves together with the outside
field, we can get solution converged as $\sim \frac {1}{r}
{e^{-mr}}$ from this new free field equation. This inter source is
confined in very small region and particle-like. The solution is
not unique, and we can separate the solution into tow parts. The
first part satisfies usual field equation and has the usual form
$\sim e^{ip^\mu x_\mu}$. Because of this, although we changed
field equation and Lagrangian, the Quantum Field Theory survived
in this model. But very different from the first part of solution,
the second part looks like $\sim \frac {1}{r}{e^{-mr}}$. We can
see this part is totally momentum and energy free, and also, we
cannot increase or decrease its' momentum and energy. This is why
we haven't found it before. The second part of field same to be
hidden behind the first part, like "Hidden Variables". \cite{bohm}

Since we changed field equation, we have to rewrite the field Lagrangian to give the
new field equation, and the second part of field along with the inter source can
contribute a negative propagator which can cancel the quadratic divergence.

In Sec. 2, we rewrite free field equation, and get convergent
solutions. We also calculate the total electromagnetic energy of a
charge in this model and we will see it is finite. In Sec. 3, we
write down the Lagrangian which can give us the new field
equation. In Sec. 4, we will see that the second part of field
contribute nothing but a negative propagator when we calculate
cross-section. In Sec.5, we discuss new description of
wave-particle duality in this model different from the statistical
description.\cite{ballentine}

\section{Convergent solution of free field equation with inter source}
By adding an inter source to free field equation, we get new field equation and we
will see the solution of the equation is converged to the source and electromagnetic
energy of a charge is finite.

The usual free field equation of Dirac, Scalar and Maxwell particles are:

\begin{eqnarray}
(\gamma^\mu \frac{1}{i}\partial_{\mu}+m)\psi (x)=0
\end{eqnarray}

\begin{eqnarray}
(-\Box + \mu^2)\phi (x)=0
\end{eqnarray}

\begin{eqnarray}
-\Box {A_\mu} (x) =0
\end{eqnarray}

the solutions of this equations are

\begin{eqnarray}
\psi (x) = \sum_{\alpha ,p} \sqrt{ \frac{m}{E_p V}} (a_p \mu_{p\alpha} e^{ip^\mu x_\mu} + b_p^+
\mu_{p \alpha}^c e^{-i p^\mu x_\mu} )
\end{eqnarray}

\begin{eqnarray}
\phi(x) = \sum_{p} \sqrt { \frac{1}{2\omega_p V}} (a_p e^{ip^\mu x_\mu} + a_p^+ e^{-ip^\mu x_\mu
})
\end{eqnarray}

\begin{eqnarray}
A_\mu (x) = \sum_{\lambda, k} \frac{ \vec{e}_k^\lambda}{ \sqrt{2\omega_k V}} (a_k^\lambda
e^{ikx} + a_k^{\lambda +} e^{-ikx})
\end{eqnarray}

Now we put an inter source and equations become

\begin{eqnarray}
(\gamma^\mu \frac{1}{i}\partial_{\mu}+m_2)\psi (x)=-f_1 (x)
\end{eqnarray}

\begin{eqnarray}
(-\Box + \mu_2^2)\phi (x)=-f_2 (x)
\end{eqnarray}

\begin{eqnarray}
(-\Box +\bar{\mu}_2^2) {A_\mu} (x) =-f_{3\mu} (x)
\end{eqnarray}

where $m_2$, $\mu_2^2$ are not necessarily equal to m, $\mu^2$,
and $\bar{\mu}_2^2$ is not necessarily equal to 0.

To separate the solutions into two parts, we rewrite the equations (8),(9) (10) as

\begin{eqnarray}
(\gamma^\mu \frac{1}{i}\partial_{\mu}+m)\psi _1(x)-(\gamma^\mu \frac{1}{i}\partial_{\mu}+m_2)\psi_2 (x)=f_1 (x)
\end{eqnarray}

\begin{eqnarray}
(-\Box + \mu^2)\phi_1 (x)-(-\Box + \mu_2^2)\phi_2 (x)=f_2 (x)
\end{eqnarray}

\begin{eqnarray}
(-\Box) {A_{1\mu}} (x)-(-\Box +\bar{\mu}_2^2) {A_{2\mu}} (x) =f_{3\mu} (x)
\end{eqnarray}

where $\psi_1(x)$, $\phi_1(x)$, $A_{1\mu} (x)$ obey the equation (1), (2), (3), and
same as (4), (5), (6). We write first part and second part with
different sign because we hope to get a negative propagator.
\\
\\
We don't know the exactly structure of source, like a string or
brane? We assume it is confined in very small region, and to such
kind of source,
 $\psi_2(x)$, $\phi_2(x)$, $A_{2\mu} (x)$ have solutions:

\begin{eqnarray}
\psi_2(x) = a_1 \tilde{\mu}(\theta) \frac{1}{r}e^{-m_2r}
\end{eqnarray}

\begin{eqnarray}
\phi_2(x) = a_2  \frac{1}{r}e^{-\mu_2r}
\end{eqnarray}

\begin{eqnarray}
A_{2\mu}(x) = a_3 \frac{1}{r}e^{- \bar{\mu}_2 r}
\end{eqnarray}

where r should be greater than the dimensions of inter source.
\\
We can see $\psi_2(x)$, $\phi_2(x)$, $A_{2\mu} (x)$ don't take
energy or momentum. $\psi_2(x)$, $\phi_2(x)$, and $A_{2\mu}(x)$
act as "manifold" of $\psi_1(x)$, $\phi_1(x)$, $A_{1\mu}(x)$. So
on the edge of $\psi_2(x)$, $\phi_2(x)$, and $A_{2\mu}(x)$, where
$\psi_2(x)$, $\phi_2(x)$, and $A_{2\mu}(x)$ approach to zero,
$\psi_1(x)$, $\phi_1(x)$, and $A_{1\mu}(x)$ should also disappear.
Let's define at r$\geq$$\Lambda_1$, $\Lambda_2$, $\Lambda_3$,
$\psi_1(x)$, $\phi_1(x)$, $A_{1\mu}(x)$ equal to zero.
\\
We write (10), (11), (12) as another way

\begin{eqnarray}
\sum_{\tilde{a}=1,2}(\gamma^\mu \frac{1}{i}\partial_{\mu}+m_{\tilde{a}})\psi _{\tilde{a}}(x)-
2(\gamma^\mu \frac{1}{i}\partial_{\mu}+m_2)\psi_2 (x)=f_1 (x)    \\
\sum_{\tilde{a}=1,2}(-\Box + \mu_{\tilde{a}}^2)\phi_{\tilde{a}} (x)-
2(-\Box + \mu_2^2)\phi_2 (x)=f_2 (x)                             \\
\sum_{\tilde{a}=1,2}(-\Box+\bar{\mu}_{\tilde{a}}^2) {A_{{\tilde{a}}\mu}} (x)-
2(-\Box +\bar{\mu}_2^2) {A_{2\mu}} (x) =f_{3\mu} (x)
\end{eqnarray}

where $m_1=m$, $\mu_1^2=\mu^2$, $\bar{\mu}_1^2=0$

From (4) and at r$\geq \Lambda_1$, $\psi_1 (x)=0$, the electric
field of a free charge e is

\begin{eqnarray*}
\vec{E}=\frac{e\vec{r}}{4\pi\varepsilon r^3} \ \ \ \ \ \ \ \   r\geq \Lambda_1 
\end{eqnarray*}
and
\begin{eqnarray}
\vec{E}=\frac{e\vec{r}}{4\pi\varepsilon \Lambda_1^3}  \ \ \ \ \ \ \ \  r<\Lambda_1
\end{eqnarray}

It's consistent with Coulomb's Law $\vec{E}=
\frac{e\vec{r}}{4\pi\varepsilon r^3}$ at r$\geq \Lambda_1$, and
$\vec{E}=0$ at r=0.

The total Electromagnetic energy of a free charge e is

\begin{eqnarray}
E= \frac{e^2}{30\pi^2\varepsilon^2}\frac{1}{\Lambda_1}
\end{eqnarray}

\section{Lagrangian}

In this part we write down the Lagrangian which can give equation
(16), (17), (18).

To free Scalar, Dirac and Maxwell fields, the Lagrangian becomes
\begin{eqnarray}
L_0 & = &
-\bar{\psi} _{\tilde{a}}(x)(\gamma^\mu \frac{1}{i}\partial_{\mu}+
m_{\tilde{a}})\psi _{\tilde{a}}(x)+
2\bar{\psi} _2(x)(\gamma^\mu \frac{1}{i}\partial_{\mu}+m_2)\psi_2 (x)  \nonumber \\
&   & +\bar{f}_1(x)\psi_2(x)+\bar{\psi}_2(x)f_1(x)   \nonumber  \\
&   &-\frac{1}{2}[\partial_{\mu}\phi_{\tilde{a}} (x)\partial^{\mu}\phi_{\tilde{a}} (x) +
\mu_{\tilde{a}}^2\phi_{\tilde{a}} (x)\phi_{\tilde{a}} (x)]
+[\partial_{\mu}\phi_2 (x)\partial^{\mu}\phi_2 (x)+ \mu_2^2\phi_2^2 (x)] \nonumber \\
&   &+f_2(x)\phi_2 (x)   \nonumber \\
&   & -\frac{1}{4}F_{\tilde{a}}^{\mu\nu}(x)F_{\tilde{a}\mu\nu}(x) \nonumber \\
&   &+\frac{1}{2}F_2^{\mu\nu}F_{2\mu\nu}+\bar{\mu}_2^2A_{2\mu}(x)A_2^{\mu}(x) \nonumber \\
&   &+f_{3\mu}(x)A_2^{\mu}(x)
\end{eqnarray}

where

\begin{eqnarray}
F_{1\mu\nu}=\partial_{\mu}A_{1\nu}-\partial_{\nu}A_{1\mu} \nonumber \\
F_{2\mu\nu}=\partial_{\mu}A_{2\nu}-\partial_{\nu}A_{2\mu}
\end{eqnarray}

Approximately, we write $f_1(x)\sim c_1\psi_2(x)$,$f_2(x)\sim c_2\phi_2(x)$, 
$f_{3\mu}(x)\sim c_3A_{2\mu}(x)$, then (21) can be written as

\begin{eqnarray}
L_0 & = &
-\bar{\psi} _{\tilde{a}}(x)(\gamma^\mu \frac{1}{i}\partial_{\mu}+
m_{\tilde{a}})\psi _{\tilde{a}}(x)+
2\bar{\psi} _2(x)(\gamma^\mu \frac{1}{i}\partial_{\mu}+m_2)\psi_2 (x)  \nonumber \\
&   &-\frac{1}{2}[\partial_{\mu}\phi_{\tilde{a}} (x)\partial^{\mu}\phi_{\tilde{a}} (x) +
\mu_{\tilde{a}}^2\phi_{\tilde{a}} (x)\phi_{\tilde{a}} (x)]
+[\partial_{\mu}\phi_2 (x)\partial^{\mu}\phi_2 (x)+ \mu_2^2\phi_2^2 (x)] \nonumber \\
&   &-\frac{1}{4}F_{\tilde{a}}^{\mu\nu}(x)F_{\tilde{a}\mu\nu}(x) \nonumber \\
&   &+\frac{1}{2}F_2^{\mu\nu}F_{2\mu\nu}+\bar{\mu}_2^2A_{2\mu}(x)A_2^{\mu}(x)
\end{eqnarray}

here we replace $ m_2\to m_2-2c_1,\mu_2^2 \to \mu_2^2-2c_2, \bar{\mu}_2^2 \to \bar{\mu}_2^2-c_3$.
\\
\\
And the interaction term of Dirac and Maxwell fields becomes

\begin{eqnarray}
L_I=\bar{\psi}_{\tilde{a}}\partial^{\mu}\psi_{\tilde{a}}A_{\tilde{a}\mu}
\end{eqnarray}
\\
The usual $U(1)\times SU(2)_L$ gauge invariant Lagrangian density for Leptons \cite{kaku} is

\begin{eqnarray}
L=L_1+L_2+L_3
\end{eqnarray}

where

\begin{eqnarray}
L_1& = &-\frac{1}{4}W_{\mu\nu}^aW^{a\mu\nu} -\frac{1}{4}F_{\mu\nu}F^{\mu\nu} \\
L_2& = &-\bar{R}\frac{1}{i}\gamma^{\mu}D_{\mu}R-\bar{L}\frac{1}{i}\gamma^{\mu}D_{\mu}L \\
L_3& = &-D_{\mu}\phi^+D^{\mu}\phi-m^2\phi^+\phi-\lambda{(\phi^+\phi)}^2 \nonumber \\
&   &-Ge(\bar{L}\phi R+\bar{R}\phi^+L)
\end{eqnarray}

where

\begin{eqnarray}
W_{\mu\nu}^a& = &\partial_{\mu} W_{\nu}^a-\partial_{\nu} W_{\mu}^a+gf^{abc}W_{\mu}^b W_{\nu}^c \\
F_{\mu\nu}& = &\partial_{\mu}B_{\nu}-\partial_{\nu}B_{\mu}  \\
D_{\mu}R& = &(\partial_{\mu}+ig^{\prime}B_{\mu})R  \\
D_{\mu}L& = &[\partial_{\mu}+\frac{i}{2}g^{\prime}B_{\mu}-\frac{i}{2}g\sigma_iW_{\mu}^i]L  \\
D_{\mu}\phi& = &[\partial_{\mu}-\frac{i}{2}g\sigma_iW_{\mu}^i-\frac{i}{2}g^{\prime}B_{\mu}]\phi
\end{eqnarray}

Now considering the dynamics of source, the total Lagrangian density for Leptons
becomes

\begin{eqnarray}
L=\tilde{L}_1+\tilde{L}_2+\tilde{L}_3
\end{eqnarray}

where

\begin{eqnarray}
\tilde{L}_1& = &-\frac{1}{4}W_{\tilde{a}\mu\nu}^aW_{\tilde{a}}^{a\mu\nu}
-\frac{1}{4}F_{\tilde{a}\mu\nu}F_{\tilde{a}}^{\mu\nu} \nonumber \\
&   &+\frac{1}{2}(\partial_{\mu} W_{2\nu}^a-\partial_{\nu} W_{2\mu}^a)
(\partial^{\mu}W_2^{a\nu}-\partial^{\nu}W_2^{a\mu}) \nonumber \\
&   &+\frac{1}{2}(\partial_{\mu} B_{2\nu}-\partial_{\nu} B_{2\mu})
(\partial^{\mu}B_2^{\nu}-\partial^{\nu}B_2^{\mu}) \nonumber \\
&   &+M_{2wab}W_{2\mu}^aW_2^{b\mu}+M_{2B}B_{2\mu}B_2^{\mu}     \\
\tilde{L}_2& = &-\bar{R}_{\tilde{a}}\frac{1}{i}\gamma^{\mu}D_{\tilde{a}\mu}R_{\tilde{a}}-\bar{L}_{\tilde{a}}\frac{1}{i}\gamma^{\mu}D_{\tilde{a}\mu}L_{\tilde{a}} \nonumber \\
&   &+2\bar{R}_2\frac{1}{i}\gamma^{\mu}\partial_{\mu}R_2+2\bar{L}_2\frac{1}{i}\gamma^{\mu}\partial_{\mu}L_2 \nonumber \\
&   &+m_{2e}(e_{2L}^+e_{2R}+e_{2R}^+e_{2L}) \nonumber \\
&   &+P(\nu_2)   \\
\tilde{L}_3& = &-D_{1\mu}\phi_1^+D_1^{\mu}\phi_1-m_1^2\phi_1^+\phi_1-\lambda{(\phi_{\tilde{a}}^+\phi_{\tilde{a}})}^2 \nonumber \\
&   &-Ge(\bar{L}_1\phi_1 R_1+\bar{R}_1\phi_1^+L_1) \nonumber \\
&   &+\partial_{\mu}\phi_2^+\partial^{\mu}\phi_2 +m_2^2\phi_2^+\phi_2
\end{eqnarray}

where

\begin{eqnarray}
W_{1\mu\nu}^a& = &\partial_{\mu} W_{1\nu}^a-\partial_{\nu} W_{1\mu}^a+gf^{abc}W_{1\mu}^b W_{1\nu}^c \\
W_{2\mu\nu}^a& = &\partial_{\mu} W_{2\nu}^a-\partial_{\nu} W_{2\mu}^a+gf^{abc}W_{2\mu}^b W_{2\nu}^c \\
F_{1\mu\nu}& = &\partial_{\mu}B_{1\nu}-\partial_{\nu}B_{1\mu}  \\
F_{2\mu\nu}& = &\partial_{\mu}B_{2\nu}-\partial_{\nu}B_{2\mu}  \\
D_{1\mu}R_1& = &(\partial_{\mu}+ig^{\prime}B_{1\mu})R_1  \\
D_{2\mu}R_2& = &(\partial_{\mu}+ig^{\prime}B_{2\mu})R_2  \\
D_{1\mu}L_1& = &[\partial_{\mu}+\frac{i}{2}g^{\prime}B_{1\mu}-\frac{i}{2}g\sigma_iW_{1\mu}^i]L_1  \\
D_{2\mu}L_2& = &[\partial_{\mu}+\frac{i}{2}g^{\prime}B_{2\mu}-\frac{i}{2}g\sigma_iW_{2\mu}^i]L_2  \\
D_{1\mu}\phi_1& = &[\partial_{\mu}-\frac{i}{2}g\sigma_iW_{1\mu}^i-\frac{i}{2}g^{\prime}B_{1\mu}]\phi_1
\end{eqnarray}
\\
where under U(1) and $SU(2)_L$ transformation, $W_2{\mu}, B_{2\mu}, R_2, L_2,\phi_2$
are invariants.
\\
\\
$P(\nu_2)$ is the term to give $\nu_2$ mass. One of possibilities is

\begin{eqnarray}
P(\nu_2)=m_{2\nu}(\bar{\nu}_{2L}\nu_{2R}+\bar{\nu}_{2R}\nu_{2L})
\end{eqnarray}

that means, although possibly $\nu_{1R}$ doesn't exist, $\nu_{2R}$ does.
\\
\\
To quarks, the usual Lagrangian density is \cite{kaku}

\begin{eqnarray}
L=-\frac{1}{4}F_{\mu\nu}^aF^{a\mu\nu}-\sum_{i=1}^6\bar{\psi}_i(\frac{1}{i}\gamma^\mu
D_{\mu}+m_i)\psi_i
\end{eqnarray}

now including the inter source, the Lagrangian becomes

\begin{eqnarray}
L& = &-\frac{1}{4}F_{\tilde{a}\mu\nu}^aF_{\tilde{a}}^{a\mu\nu}-\sum_{i=1}^6\bar{\psi}_{\tilde{a}i}(\frac{1}{i}\gamma^\mu
D_{\tilde{a}\mu}+m_{\tilde{a}i})\psi_{\tilde{a}i} \nonumber \\
&   &+\frac{1}{2}(\partial_{\mu}A_{2\nu}^a-\partial_{\nu}A_{2\mu}^a)
(\partial^{\mu}A_2^{a\nu}-\partial^{\nu}A_2^{a\mu})  \nonumber \\
&   &+\bar{m}_2^2A_{2\mu}^aA_2^{a\mu}  \nonumber \\
&   &+2\sum_{i=1}^6\bar{\psi}_{2i}(\frac{1}{i}\gamma^\mu
\partial_{\mu}+m_{2i})\psi_{2i}
\end{eqnarray}

\section{S-Matrix}

In Section 2, we separate Scalar,Dirac, Maxwell field into two parts (4),(5),(6) and
(13),(14),(15). The first part, (4),(5), (6), like $\sim e^{ip^{\mu}x_{\mu}}$, is
same as usual free quantum field. And the second part, (13),(14),(15), like $\sim\frac
{1}{r}e^{-mr}$, doesn't take energy and momentum. Because of this property, the
second part will contribute nothing but an negative propagator when we calculate cross-section.
\\
\\
Because of Inter-source, the S-Matrix can be written as

\begin{eqnarray}
S & = & <\psi_{\tilde{a}1}(x_1)\cdots\bar{\psi}_{\tilde{a}1}(y_1)\cdots\phi_{\tilde{a}1}(z_1)\cdots
|\psi_{\tilde{a}1}^{\prime}(x_1^{\prime})\cdots\bar{\psi}_{\tilde{a}1}^{\prime}(y_1^{\prime})\cdots\phi_{\tilde{a}1}^{\prime}(z_1^{\prime})\cdots>
  \nonumber \\
& = & \int{dx_1}\cdots{dx_1^{\prime}}\cdots{dy_1}\cdots{dy_1^{\prime}}\cdots{dz_1}\cdots{dz_1^{\prime}}\cdots
 <\psi_{\tilde{a}1}(x_1)\cdots\bar{\psi}_{\tilde{a}1}(y_1)\cdots\phi_{\tilde{a}1}(z_1)\cdots|  \nonumber \\
&   & \psi_{\tilde{a}1}(x_1)\cdots\bar{\psi}_{\tilde{a}1}(y_1)\cdots\phi_{\tilde{a}1}(z_1)\cdots
R\psi_{\tilde{a}1}^{\prime}(x_1^{\prime})\cdots\bar{\psi}_{\tilde{a}1}^{\prime}(y_1^{\prime})\cdots\phi_{\tilde{a}1}^{\prime}(z_1^{\prime})\cdots \nonumber \\
&  & |\psi_{\tilde{a}1}^{\prime}(x_1^{\prime})\cdots\bar{\psi}_{\tilde{a}1}^{\prime}(y_1^{\prime})\cdots\phi_{\tilde{a}1}^{\prime}(z_1^{\prime})\cdots>
\end{eqnarray}

But because of the Inter-source, $|\psi_2>, |\phi_2>,|A_{2\mu}> $ cannot take energy
or momentum, and also we cannot increase or decrease their energy or momentum,
so we can write

\begin{eqnarray}
\psi_{\tilde{a}}|\psi_{\tilde{a}}> & = & \psi_1|\psi_1>+\psi_2|\psi_2> \nonumber \\
& = & \psi_1|\psi_1>  \\
\phi_{\tilde{a}}|\phi_{\tilde{a}}> & = & \phi_1|\phi_1>+\phi_2|\phi_2>  \nonumber  \\
& = & \phi_1|\phi_1>  \\
A_{\tilde{a}\mu}|A_{\tilde{a}\mu}>& = & A_{1\mu}|A_{1\mu}>+A_{2\mu}|A_{2\mu}> \nonumber \\
& = & A_{1\mu}|A_{1\mu}>
\end{eqnarray}

then S-Matrix becomes

\begin{eqnarray}
S & = & \int{dx_1}\cdots{dx_1^{\prime}}\cdots{dy_1}\cdots{dy_1^{\prime}}\cdots{dz_1}\cdots{dz_1^{\prime}}\cdots
 <\psi_{11}(x_1)\cdots\bar{\psi}_{11}(y_1)\cdots\phi_{11}(z_1)\cdots|
\psi_{11}(x_1)\cdots{\psi}_{11}(y_1)\cdots\phi_{11}(z_1)\cdots \nonumber \\
&   & R\psi_{11}^{\prime}(x_1^{\prime})\cdots\bar{\psi}_{11}^{}\prime(y_1^{\prime})\cdots\phi_{11}^{\prime}(z_1^{\prime})\cdots
|\psi_{11}^{\prime}(x_1^{\prime})\cdots{\psi}_{11}^{\prime}(y_1^{\prime})\cdots\phi_{11}^{\prime}(z_1^{\prime})\cdots>
\end{eqnarray}

But $\psi_2, \phi_2, A_{2\mu} $ can contribute an negative propagator to S-Matrix.

\begin{eqnarray}
S(x,y)& = &i<0|T(\bar{\psi}_{\tilde{a}}(x)\psi_{\tilde{a}}(y))|0> \nonumber \\
& = & i<0|T(\bar{\psi}_1(x)\psi_1(y))|0>+i<0|T(\bar{\psi}_2(x)\psi_2(y))|0>  \\
\Delta(x,y) & = &i<0|T(\phi_{\tilde{a}}(x)\phi_{\tilde{a}}(y))|0> \nonumber \\
& = & i<0|T(\phi_1(x)\phi_1(y))|0>+i<0|T(\phi_2(x)\phi_2(y))|0>  \\
D_{\mu\nu}(x,y) & = &i<0|T(A_{\tilde{a}\mu}(x)A_{\tilde{a}\nu}(y))|0> \nonumber \\
& = &i<0|T(A_{1\mu}(x)A_{1\nu}(y))|0>+i<0|T(A_{2\mu}(x)A_{2\nu}(y))|0>
\end{eqnarray}

From the Lagrangian(23), to first order, we can get

\begin{eqnarray}
S^0(p)& = &\frac{1}{\gamma p+m} -\frac{1}{\gamma p+m_2}  \\
\Delta^0(p)& = &\frac{1}{p^2+\mu^2} -\frac{1}{p^2+\mu_2^2}  \\
D_{\mu\nu}^0(k)& = &\frac{\eta_{\mu\nu}}{k^2}-\frac{\eta_{\mu\nu}}{k^2+\bar{\mu}_2^2}
\end{eqnarray}

where $m_2,\mu_2^2,\bar{\mu}_2^2$ should be very big.

When we calculate tree graph, because $m_2,\mu_2^2,\bar{\mu}_2^2$
are very big, the second term of (58),(59),(60) are very small,
but when we calculate loop graph, they can cancel quadratic
divergence.
\\
\\
In the same way we can write the propagator of gauge field

\begin{eqnarray}
D_{\mu\nu}^{\alpha\beta}(x,y) & = &i<0|T(A_{\tilde{a}\mu}^{\alpha}(x)A_{\tilde{a}\nu}^{\beta}(y))|0> \nonumber \\
& = &i<0|T(A_{1\mu}^{\alpha}(x)A_{1\nu}^{\beta}(y))|0>+i<0|T(A_{2\mu}^{\alpha}(x)A_{2\nu}^{\beta}(y))|0>
\end{eqnarray}

From Lagrangian (49), to first order and Feynman gauge, the
propagator becomes
\begin{eqnarray}
D_{\mu\nu}^{0\alpha\beta}(k) =\frac{\eta_{\mu\nu}}{k^2}\delta^{\alpha\beta}-\frac{\eta_{\mu\nu}}{k^2+\bar{m}^2}\delta^{\alpha\beta}
\end{eqnarray}

where $\bar{m}^2$ should also be very big.

\section{Wave-Particle Duality}

We assume every real particle has an inter source. Because the
source is confined in a very small region, as assumed, particle
has an inter core which make it particle-like. On the other hand,
the quantum field, outside its inter source, can move like wave
and make it wavelike. So, here, we give a new description of
wave-particle duality.

{}

\begin{thebibliography}{99}
\bibitem{bohm} D.Bohm, A suggested Interpretation of the Quantum Theory in terms of
 "Hidden" Variables.I,II.Phys.Rev.85,166-193(1953) \\
J.B.Keller, Bohm's Interpretation of the Quantum Theoryin terms of "Hidden" Variables.
Phys.Rev.89,1040-1041(1953).
\bibitem{ballentine} L.E.Ballentine, Statistical Interpretation of Quantum Mechanics.Rev.Mod.
Phys.42,358-381(1970).
\bibitem{kaku} M.Kaku, Quantum Field Theory (1993).
\end{thebibliography}
\end{document}